\documentclass[amsmath,prb,twocolumn,showpacs,superscriptaddress]{revtex4}

\usepackage{color}
\usepackage{amsfonts}
\usepackage{graphicx}
\usepackage{pstricks}

\begin{document}

\title{Transport and optical response of molecular junctions driven by surface plasmon-polaritons}
\date{\today}
\author{Maxim Sukharev}
\email{maxim.sukharev@asu.edu}
\affiliation{Department of Applied Sciences and Mathematics, 
Arizona State University at the Polytechnic Campus, Mesa, AZ 85212, USA}
\author{Michael Galperin}
\email{migalperin@ucsd.edu}
\affiliation{Department of Chemistry \& Biochemistry, University of California 
at San Diego, La Jolla, CA 92093, USA}
\begin{abstract}
We consider a biased molecular junction subjected to external time-dependent
electromagnetic field. The field for two typical junction 
geometries (bowtie antennas and metal nanospheres) is calculated within 
finite-difference time-domain technique. 
Time-dependent transport and optical response of
the junctions is calculated within non-equilibrium Green's function approach
expressed in a form convenient for description of multi-level systems.
We present numerical results for a two-level (HOMO-LUMO) model, 
and discuss influence of localized surface plasmon polariton modes 
on transport.
\end{abstract}


\maketitle
\section{Introduction}\label{intro}
Optical properties of structures composed of noble metals have long been 
attracting a considerable attention due to unique features of such systems 
in the visible spectrum.\cite{Atwater98,Ozbay,Barnes07,Zhang08} 
Recent advances in fabrication techniques\cite{Fendler04} along with 
a tremendous progress in laser technologies opened new venues for application 
of plasmonic materials in biology,\cite{VanDuyne04} 
integrated optics,\cite{Berini05} nanoscale imagining,\cite{ShalaevBook07} 
and single molecule manipulation.\cite{Yanagida98} 
Physics of surface plasmon phenomenon is relatively simple and has long been 
studied.\cite{RaetherBook88,VollmerBook95} In brief, coherent oscillations 
of conductive electrons in a skin-layer of metal known as plasmons are 
capable of producing strong local electromagnetic (EM) fields in the 
near-field region. It has been reported that such "hot" spots can be 
localized within $10$~nm or less. This along with a great sensitivity 
to initial conditions and geometry makes plasmonic structures so attractable 
for atom/molecule manipulations.

A natural combination of nanoplasmonics and molecular response 
to the generated field started to appear as molecular 
nanopolaritonics,\cite{LopataNeuhauser,LopataNeuhauser2} 
which studies molecular influence on field propagation, 
and as a tool for developing molecular switches.\cite{SukharevSeideman} 
The latter utilizes nonadiabatic alignment of a molecule on semiconductor 
surface under a tip of scanning tunneling microscope.

Recent developments in experimental techniques capable of measuring optical 
response of current-carrying molecular junctions\cite{Ho,Natelson}
lead to theoretical formulations suitable for simultaneous 
description of both transport and optical properties of molecular 
devices.\cite{opt,raman} 

While experimental data are measured in real time,
theoretical description of both transport and optical response so far 
has mostly been focused on a steady-state description.
Time-dependent transport usually is treated either within 
kinetic theory\cite{Petrov2005,Petrov2006} or within time-dependent 
density functional approach.\cite{Car,Almbladh,Gross}
The former generally misses broadening of molecular states
due to coupling to macroscopic contacts\cite{Wacker05,Neuhauser,Esposito} and 
information on coherence,\cite{vonOppen} although interesting generalizations
started to appear.\cite{Wegewijs}
Limitations of the latter are due to absence of developed pseudopotentials
and fundamental necessity to treat finite (closed) systems (see e.g.
Ref.~\onlinecite{Tretiak} for discussion). An alternative approach, based on
non-equilibrium Green function (NEGF) technique, was initially 
formulated in Refs.~\onlinecite{JauhoWingreenMeir,AnantramDatta,Guo}. 
This approach is a natural choice
for description of open non-equilibrium systems. Moreover it provides 
possibility to describe response of a molecular junction initially under bias
to external time-dependent perturbation (e.g. laser field).

Here we consider influence of external field specific for particular
geometry on transport properties and optical response of
molecular junction. While formulation of time-dependent transport
within NEGF is general,\cite{JauhoWingreenMeir,AnantramDatta}
all the applications so far were restricted to resonant single level
models only. We propose a variant of the scheme capable of dealing with
many-level systems. The exact calculations are compared to adiabatic
pumping regime, frequent in the literature on time-dependent 
transport,\cite{Dunietz,Mucciolo} were at the lowest order 
the problem is reduced to a set of quasi-steady-state 
solutions with time dependent (slow timescale) parameters.
Also we generalize our previous consideration of steady-state optical 
response of current-carrying junctions\cite{Nitzan_prl,Nitzan_jcp} 
to a time-dependent situation.  

The paper is organized as follows. Section \ref{model} presents a model 
of molecular junction.
Section \ref{fdtd} describes methodology of EM field calculation.
Section \ref{td} describes methodology for simulating transport through 
molecular junction subjected to external time-dependent field. 
Adiabatic pumping version is discussed in section \ref{atd}.
Numerical results are presented in section \ref{numres}.
Section \ref{conclude} concludes.

\section{\label{model}Model}
We consider a two-level system $\varepsilon_{1,2}$, representing
highest occupied (HOMO) and lowest unoccupied (LUMO)
molecular orbitals (or ground and excited states in the many-body
language), coupled to two macroscopic electrodes $L$ and $R$.
The electrodes are considered to be each in its own equilibrium with 
electrochemical potentials $\mu_L$ and $\mu_R$, respectively.
We assume that the driving (laser field) frequency is smaller than the 
plasma frequency, so that usual division of the junction into non-equilibrium
molecule coupled to free electron reservoirs (metallic contacts) is 
relevant (for a thorough discussion of the assumptions see 
Ref.~\onlinecite{JauhoWingreenMeir}).   
Local field at the position of the molecule is calculated within 
finite difference time domain technique (see section~\ref{fdtd} for details),
and is assumed to be an external time-dependent driving force 
causing (de)excitation in the molecule. Following Ref.~\onlinecite{Nitzan_jcp}
in addition to charge transfer between contacts and molecule we introduce also
energy transfer (coupling of molecular excitations to electron-hole excitations
in the contacts). Molecular excitations are 
coupled to a bath of free photon modes (accepting modes), which serve as a
measurement device of molecular optical response.
Hamiltonian of the system is
\begin{align}
 \label{H}
 \hat H =& \hat H_0 + \hat V
 \\
 \label{H0}
 \hat H_0 =& \sum_{i=1,2}\varepsilon_{i}\hat d_i^\dagger\hat d_i
          -  \left(\vec\mu_{12}\hat d_1^\dagger\hat d_2
                  +\vec\mu_{21}\hat d_2^\dagger\hat d_1\right)\vec E(t)
 \nonumber \\
          +& \sum_{k\in\{L,R\}}\varepsilon_k \hat c_k^\dagger\hat c_k
          +  \sum_\alpha \omega_\alpha \hat a_\alpha^\dagger\hat a_\alpha
 \\
          +& \sum_{i=1,2;k\in\{L,R\}}\left(V_{ki}^{et}\hat c_k^\dagger\hat d_i
                                 +V_{ik}^{et}\hat d_i^\dagger\hat c_k\right)
 \nonumber \\
 \label{V}
 \hat V =& \sum_{k\neq k'\in\{L,R\}}
        \left(V_{kk'}^{eh}\hat c_k^\dagger\hat c_{k'}\hat d_2^\dagger\hat d_1
             +V_{k'k}^{eh}\hat c_{k'}^\dagger\hat c_k\hat d_1^\dagger\hat d_2
        \right)
        \\
        +& \qquad 
        \sum_\alpha\left(V_\alpha^{p}\hat a_\alpha\hat d_2^\dagger\hat d_1
        +\overset{*}{V}{}_\alpha^{p}\hat a_\alpha^\dagger\hat d_1^\dagger\hat d_2
        \right)
        \nonumber
\end{align}
Here $\hat d_i^\dagger$ ($\hat d_i$) and $\hat c_k^\dagger$ ($\hat c_k$)
are creation (annihilation) operators for an electron in the state $i$ of 
the molecule and state $k$ of the contact, respectively. 
$\hat a_\alpha^\dagger$ ($\hat a_\alpha$) is creation (annihilation) operator 
for a photon in the state $\alpha$, $\vec E(t)$ is external time-dependent
field, and $\vec\mu_{ij}=<i|\hat{\vec\mu}|j>$ is matrix element of 
the molecular (vector) dipole operator between states $i$ and $j$ of the 
molecule ($i,j=1,2$). We assume $\vec\mu_{11}=\vec\mu_{22}=0$ (or alternatively
one can think about these contributions being included into definition of
the state energies $\varepsilon_{1,2}$). $V^{et}$ and $V^{en}$
are matrix elements for electron and energy transfer between
molecule and contacts, and $V^{p}$ represents optical response of the molecule.

Below we consider two approaches to transport and optical response simulations
within the model: exact solution of the time-dependent Dyson equation
and adiabatic pumping regime. The former is similar to the procedure described 
in Refs.~\onlinecite{JauhoWingreenMeir,AnantramDatta,Guo},
however it is presented in a form convenient for treating a multi-level 
molecular system (see section~\ref{td} for discussion).
The latter assumes that $\vec E(t)$ can be represented as a product
of an oscillation of frequency $\omega_0$ with a slowly varying in time
(on the timescale of $\omega_0$) envelope $\vec F(t)$. 
In the spirit of the Born-Oppenheimer approximation $F(t)$ is considered as 
a parameter when solving electronic part of the problem. 
In this case the form of molecule-field 
interaction becomes (within rotating wave approximation)
\begin{equation}
 \label{adiab}
 -\left(\vec\mu_{12}\hat d_1^\dagger\hat d_2 e^{i\omega_0 t}
       +\vec\mu_{21}\hat d_2^\dagger\hat d_1 e^{-i\omega_0 t}
  \right)\vec F(t)
\end{equation}
Details of the approach are presented in section~\ref{atd}.

As usual, we treat the perturbation $\hat V$, Eq.(\ref{V}), at the second
order and within noncrossing approximation.\cite{Mahan} 
Self-energy due to energy transfer (on the Keldysh contour) is\cite{Nitzan_jcp}
\begin{align}
 \label{Sen}
 \Sigma^{en}(\tau_1,\tau_2) =& \sum_{k\neq k'\in\{L,R\}} |V_{kk'}|^2
 g_k(\tau_2,\tau_1) g_{k'}(\tau_1,\tau_2)
 \nonumber \\
 \times & \left[
 \begin{array}{cc}
  G_{22}(\tau_1,\tau_2) & G_{21}(\tau_1,\tau_2) \\
  G_{12}(\tau_1,\tau_2) & G_{11}(\tau_1,\tau_2)
 \end{array}
 \right]
\end{align}
where $G_{ij}$ are molecular Green functions in the lowest order of 
expansion associated with the Hamiltonian $\hat H_0$, Eq.(\ref{H0}),
and $g_k$ are Green functions of free electrons in the contacts.
Self-energy due to coupling to photon bath is\cite{Nitzan_jcp}
\begin{widetext}
\begin{equation}
 \label{Sp}
 \Sigma^{p}(\tau_1,\tau_2) = \sum_\alpha |V_\alpha^{p}|^2
 \left[
 \begin{array}{cc}
 iF_\alpha(\tau_2,\tau_1)G_{22}(\tau_1,\tau_2) &
 \delta(\tau_1,\tau_2)\int_{-\infty}^{t_1} dt'\,\rho_{12}(t') F_\alpha^a(t'-t_1)
 \\
 \delta(\tau_1,\tau_2)\int_{-\infty}^{t_2} dt'\,F_\alpha^r(t_1-t')\rho_{21}(t')
 & iF_\alpha(\tau_1,\tau_2)G_{11}(\tau_1,\tau_2)
 \end{array}
 \right]
\end{equation}
\end{widetext}
where $F_\alpha$ is Green function for free photon and
$\rho_{ij}(t)\equiv -iG^{<}_{ij}(t,t)$ is non-equilibrium reduced 
density matrix.

Below we discuss methods for calculating external field for different 
geometries, and present approaches to calculate time-dependent 
current and optical response of driven molecular junction. 

\section{\label{fdtd}Electromagnetic field simulations}
Among various numerical techniques that allow one to predict optical 
properties of plasmonic systems the finite-difference time-domain approach 
(FDTD) is considered to be the most efficient and yet relatively simple. 
FDTD yields data in perfect agreement with experimental measurements and 
results obtained within other techniques.\cite{FDTDbenchmarks} 
We simulate optical response of metal structures utilizing FDTD approach, 
in which Maxwell equations are discretized in space and time following 
Yee's algorithm\cite{TafloveBook}. Dispersion of dielectric constant of 
metal, $\varepsilon(\omega)$, is taken in the form of the Drude model
\begin{align}
\label{Drude model}
    &\varepsilon(\omega)=\varepsilon_r-\frac{\omega^{2}_{p}}{\omega^{2}-i \Gamma \omega}
    &  
\end{align}
with numerical parameters describing silver for the wavelengths of interest 
$\varepsilon_r=8.26$, $\omega_p=1.76\times10^{16}$ rad/sec, 
$\Gamma=3.08\times10^{14}$ rad/sec.

For simulations of open systems, one needs to impose artificial absorbing 
boundaries in order to avoid reflection of outgoing EM waves back 
to the simulation domain.  Among various approaches that address this 
numerical issue, the perfectly matched layers (PML) technique\cite{BerengerPML} 
is considered to be the most adequate. It reduces the reflection coefficient 
of outgoing waves at the simulation region boundary to $~10^{-8}$.  
Essentially, the PML approach surrounds the simulation domain by thin layers 
of non-physical material that efficiently absorbs outgoing waves incident 
at any angle.  We implement the most efficient and least memory intensive 
method, convolution perfectly matched layers (CPML)\cite{CPML} absorbing 
boundaries, at all six sides of the 3D modeling space. Through extensive 
numerical experimentation, we have empirically determined optimal parameters 
for the CPML boundaries that lead to almost no reflection of the outgoing 
EM waves at all incident angles. Spatial steps, $\delta x=\delta y=\delta z$, 
along all axes are fixed at $1$~nm to assure numerical convergence and 
the temporal step is $\delta t = \delta x/(2 c)$, where $c$ is the speed of 
light in vacuum.

Numerical integration of Maxwell equations on a grid within the FDTD 
framework was performed at the local ASU home-built supercomputer 
utilizing 120 processors. An average execution time for our codes is around 
20 minutes.

A particular advantage of the FDTD method is its ability to obtain 
the optical response of the structure (assuming linear response) in 
the desired spectral range in a single run.\cite{SukharevSeideman07}  
The system is excited with an ultra-short optical pulse constructed from 
Fourier components spanning the frequency range of interest.  
Next, Maxwell's equations are propagated in time for several hundred 
femtoseconds and the components of the EM field are detected at the point of 
interest (for our purposes we consider the detection point where a molecule 
is located).  Fourier transforming the detected EM field on the fly yields 
intensities that can be easily processed into the spectral response.  
Since we also have access to the field components, we can evaluate 
the intensity enhancement relative to the incident field. This provides 
the capability for straightforward evaluation of `coupling efficiency' of our 
plasmonic structures in the spectral range of interest.

\section{\label{td}Time-dependent transport}
We are interested in calculating time-dependent current and optical response of
the junction. Expression for the current at the interface $K$ 
($K=L,R$) between molecule and contact is\cite{HaugJauho}
\begin{align}
 \label{IKt}
 &I_K(t) = \frac{e}{\hbar}\int_{-\infty}^t dt_1 
 \nonumber \\
 &\mbox{Tr}
 \left[\mathbf{\Sigma}_K^{<}(t,t_1)\,\mathbf{G}^{>}(t_1,t) 
      +\mathbf{G}^{>}(t,t_1)\,\mathbf{\Sigma}_K^{<}(t_1,t)
 \right. \\ & \left. \ \
      -\mathbf{\Sigma}_K^{>}(t,t_1)\,\mathbf{G}^{<}(t_1,t)
      -\mathbf{G}^{<}(t,t_1)\,\mathbf{\Sigma}_K(t_1,t)
 \right]
 \nonumber
\end{align}
where $\Sigma_K$ is self-energy due to coupling to contact $K$
\begin{equation}
 \label{Set}
 \left[\mathbf{\Sigma}^{et}_K(\tau_1,\tau_2)\right]_{ij}=\sum_{k\in K}
 V_{ik} g_k(\tau_1,\tau_2) V_{kj} 
\end{equation}
and $r$, $a$, $<$, $>$ are retarded, advanced, lesser, and greater
projections respectively. In the wide band limit, when escape rate
matrix 
\begin{equation}
 \left[\mathbf{\Gamma}_K(E)\right]_{ij} = 
 2\pi \sum_{k\in K}V_{ik}V_{kj}\delta(E-\varepsilon_k)
\end{equation}
is assumed to be energy independent and real part of the self-energy 
(\ref{Set}) is disregarded, and when time modulation is restricted to molecular
subspace only, expression (\ref{IKt}) can be reduced
to\cite{JauhoWingreenMeir}
\begin{align}
 \label{IKtwbl}
 I_K(t) =& I_K^{in}(t) - I_K^{out}(t)
 \\
 \label{IKin}
 I_K^{in}(t)=&-\frac{e}{\pi\hbar}\int_{-\infty}^{+\infty}dE\, f_K(E)
  \mbox{Im}\mbox{Tr}\left[\mathbf{\Gamma}_K\mathbf{A^r}(t,E)\right]
 \\
 \label{IKout}
 I_K^{out}(t) =&+\frac{e}{\hbar}
 \mbox{Re}\mbox{Tr}\left[\mathbf{\Gamma}_K\mathbf{\rho}(t)\right]
\end{align}
where $f_K(E)$ is Fermi-Dirac distribution in contact $K$ and
$\mathbf{A}^r(t,E)$ is time-dependent (one-sided) Fourier transform of 
the retarded Green function $\mathbf{G}^r(t,t')$. 
\begin{equation}
 \label{Ar}
 \mathbf{A}^r(t,E) = \int_{-\infty}^t dt' e^{iE(t-t')} \mathbf{G}^r(t,t')
\end{equation}
In the absence of time-dependent driving 
$\mathbf{A}^r(t,E)$ reduces to usual Fourier transform for retarded Green 
function $\mathbf{G}_0^r(E)=[E-\mathbf{H}_0-\mathbf{\Sigma}^{r}(E)]^{-1}$.
In general $\mathbf{\Sigma}^r$ has contributions (additive within
noncrossing approximation) from all the processes involved.
$\mathbf{\rho}(t)$ in (\ref{IKout}) is reduced density matrix
\begin{equation}
 \label{rho}
 \mathbf{\rho}(t) = -i\mathbf{G}^{<}(t,t)
\end{equation}
Lesser and greater Green functions are calculated
from the time dependent Dyson equation
\begin{align}
 \mathbf{G}^{>,<}(t,t') =& \int_{-\infty}^{t} dt_1\int_{-\infty}^{t'} dt_2
 e^{-iE(t_1-t_2)}
 \nonumber \\ \times &
 \mathbf{A}^r(t_1,E)\,\mathbf{\Sigma}^{>,<}(E)\,\mathbf{A}^a(t_2,E)
\end{align}
where 
\begin{equation}
 \label{Aa}
 A^a_{ij}(t,E)=\overset{*}{A}{}^r_{ji}(t,E)
\end{equation}
and $\mathbf{A}^r(t,E)$ is defined in Eq.(\ref{Ar}).

Contrary to our previous consideration\cite{Nitzan_prl,Nitzan_jcp}
optical response of molecular junction is calculated as a true photon flux
into modes $\{\alpha\}$, rather than corresponding electronic current
between molecular orbitals. We start from general expression for
time-dependent photon flux into mode $\alpha$ (the derivation follows 
the corresponding procedure for electronic current, the latter 
can be found in e.g. Ref.~\onlinecite{HaugJauho})
\begin{align}
 \label{Jat}
 J_\alpha(t) \equiv & \frac{d}{dt}<\hat a_\alpha^\dagger(t)\hat a_\alpha(t)>
 = |V_\alpha^{p}|^2 \int_{-\infty}^t dt_1
 \nonumber \\
 \times& \left[
    F_{\alpha}^{<}(t,t_1)\mathcal{G}^{>}(t_1,t) + 
    \mathcal{G}^{>}(t,t_1)F_{\alpha}^{<}(t_1,t) 
 \right. \\ -& \left. \,
    F_{\alpha}^{>}(t,t_1)\mathcal{G}^{<}(t_1,t) -
    \mathcal{G}^{<}(t,t_1)F_{\alpha}^{>}(t_1,t)
    \right]
 \nonumber
\end{align}
Here $\mathcal{G}$ is two-particle Green function
\begin{equation}
 \mathcal{G}(\tau,\tau') \equiv -\frac{i}{\hbar}
       <T_c \hat D(\tau)\,\hat D^\dagger(\tau')>
\end{equation}
where $\hat D\equiv \hat d_1^\dagger\hat d_2$ is molecular de-excitation 
operator. For empty accepting mode $\alpha$ expression (\ref{Jat})
reduces to
\begin{equation}
 \label{Jatacc}
 J_\alpha(t) = - 2 \frac{|V_\alpha^{p}|^2}{\hbar} 
 \mbox{Im} \int_{-\infty}^{t} dt_1\,
 e^{i\omega_\alpha(t_1-t)} \mathcal{G}^{<}(t_1,t)
\end{equation}
As in Ref.~\onlinecite{Nitzan_jcp} we approximate the two-particle Green 
function by zero-order (in interaction) expression
\begin{equation}
 \label{GF2approx}
 \mathcal{G}^{<}(t_1,t)\approx -i\hbar\left[
 G^{>}_{11}(t,t_1)G^{<}_{22}(t_1,t)-\rho_{12}(t)\rho_{21}(t_1)\right]
\end{equation}
Note that if envelope change in time is slow 
(on the timescale of $\omega_\alpha$)
second term on the right of (\ref{GF2approx}) can be safely disregarded.
In this case expression (\ref{Jatacc}) becomes equivalent to approximate
expression used in Ref.~\onlinecite{Nitzan_jcp}.

Below we calculate frequency resolved 
\begin{align}
 \label{Jwa}
 J(\omega,t) \equiv &
 \sum_{\alpha} J_\alpha(t) \delta(\omega-\omega_\alpha) 
 \\
 \approx & \frac{1}{\pi\hbar} \gamma_\alpha(\omega) 
 \mbox{Re} \int_{-\infty}^{t} dt_1
 e^{i\omega(t_1-t)} G_{11}^{>}(t,t_1)\, G_{22}^{<}(t_1,t)
 \nonumber 
\end{align}
and total 
\begin{equation}
 \label{Jtot}
 J_{tot}(t) \equiv \int_0^{\infty} d\omega J(\omega,t)
\end{equation}
photon fluxes. 
Here $\gamma_\alpha(\omega)\equiv 2\pi\sum_\alpha \delta(\omega-\omega_\alpha)$,and in simulations we use\cite{Nitzanbook}
\begin{equation}
 \label{gamma}
 \gamma_\alpha(\omega) = \eta\omega e^{-\omega/\omega_c}
\end{equation}

To calculate time-dependent charge, Eq.(\ref{IKtwbl}), and
photon, Eq.(\ref{Jatacc}), fluxes one needs time-dependent
Fourier transform of retarded Green function, Eq.(\ref{Ar}).
The Dyson equation for retarded Green function is
\begin{align}
 \label{DysG}
 &\left(i\frac{\partial}{\partial t}-\mathbf{H}(t)\right)\mathbf{G}^r(t,t')
 \\ &\quad
 -\int_{-\infty}^{+\infty}dt_1\,\mathbf{\Sigma}^r(t-t_1)\,
  \mathbf{G}^r(t_1,t') = \delta(t-t')
 \nonumber
\end{align}
Its one-sided Fourier transform leads to equation for $\mathbf{A}^r(t,E)$
in the form
\begin{align}
 \label{DysA}
 &\left(i\frac{\partial}{\partial t}-[\mathbf{H}_0(t)-E]\right)\mathbf{A}^r(t,E)
 \\ &\quad
 -\int_{-\infty}^{+\infty}dt_1\,\mathbf{\Sigma}^r(t-t_1)\,
  \mathbf{A}^r(t_1,E) = \mathbf{I}
 \nonumber
\end{align}
We consider situation when time-dependent external field is applied 
at time $t_0$ to a biased molecular junction initially at steady-state.
In this case differential equation (\ref{DysA}) 
can be solved numerically starting from known initial condition 
$\mathbf{A}^r(t_0,E)=\mathbf{G}^r_0(E)=[E-\mathbf{H}_0^c-\mathbf{\Sigma}^r(E)]^{-1}$.

Alternatively, splitting $\mathbf{H}_0(t)$ into time-independent
$\mathbf{H}_0^{c}$ and time-dependent $\mathbf{H}_0^{t}(t)$ parts
(average over time of the time-dependent part can be included into 
the time-independent Hamiltonian), one can rewrite Dyson equation (\ref{DysG})
in the integral form
\begin{equation}
 \label{DysGint}
 \mathbf{G}^r(t,t') = \mathbf{G}^r_0(t-t') + \int_{-\infty}^t dt_1\,
 \mathbf{G}^r_0(t-t_1)\mathbf{H}_0^{t}(t_1)\mathbf{G}^r(t_1,t')
\end{equation}
One-sided Fourier transform of (\ref{DysGint}) leads to integral
equation for $\mathbf{A}^r(t,E)$
\begin{align}
 \label{DysAint}
 &\mathbf{A}^r(t,E) = \mathbf{G}^r_0(E) 
 \\ & \quad
 + \int_{t_0}^{t} dt_1\,
 \mathbf{G}^r_0(t-t_1) e^{iE(t-t_1)}\mathbf{H}_0^c(t_1) 
 \mathbf{A}^r(t_1,E)
 \nonumber
\end{align}
where lower limit of the integral in the right is set to $t_0$ since
$\mathbf{H}_0^t(t<t_0)=0$. Its solution is
\begin{align}
 \label{Asolve}
 \mathbf{A}^r(t,E) =& \mathbf{U}_{eff}(t,t_0;E)\,\mathbf{G}^r_0(E)
 \\
 \label{Ueff}
 \mathbf{U}_{eff}(t,t_0;E) \equiv & T\exp\left[\int_{t_0}^t dt_1\,
 \mathbf{G}^r_0(t-t_1)e^{iE(t-t_1)}\mathbf{H}_0^t(t_1)\right]
\end{align}
Effective evolution operator $\mathbf{U}_{eff}$ can be obtained by
variety of methods available in the literature (see e.g. 
Ref.~\onlinecite{Nauts} and references therein). 
One of the simplest schemes
is cumulant (or Magnus) expansion.\cite{Magnus,Mukamel,Lamberti}

Note that although our consideration is restricted to the case when
time-dependent driving takes place in the molecular subspace only,
generalization to driving in the contacts or at the molecule-contact 
interface is straightforward.

\section{\label{atd}Adiabatic pumping regime}
When time evolution of an envelope $\vec F(t)$, Eq.(\ref{adiab}),
is slow on the timescale of the field frequency $\omega_0$, consideration of
the time dependent transport is simplified by invoking adiabatic assumption
(treating $\vec F(t)$ as a parameter). 

We start with Hamiltonian (\ref{H}) in which interaction with driving field 
is written in the form presented in Eq.(\ref{adiab}). Transforming the 
Hamiltonian into rotating frame of the field\cite{ZhagXuXie_rf,FranssonZhu_rf}
\begin{align}
 \hat{\bar H} =& e^{\hat S}\hat H e^{-\hat S} + 
 \left(i\frac{\partial}{\partial t} e^{\hat S}\right) e^{-\hat S}
 \\
 \hat S =& -\frac{i\omega_0 t}{2}\left(\hat n_1-\hat n_2\right)
\end{align}
where $\hat n_i=\hat d_i^\dagger\hat d_i$ ($i=1,2$), leads to
\begin{align}
 \label{Hbar}
 \hat{\bar H} =& \hat{\bar H}_0 + \hat{\bar V}
 \\
 \label{H0bar}
 \hat{\bar H}_0 =& \sum_{i=1,2}\bar\varepsilon_{i}\hat d_i^\dagger\hat d_i
          -  \left(\vec\mu_{12}\hat d_1^\dagger\hat d_2
                  +\vec\mu_{21}\hat d_2^\dagger\hat d_1\right)\vec F(t)
 \nonumber \\
          +& \sum_{k\in\{L,R\}}\varepsilon_k \hat c_k^\dagger\hat c_k
          +  \sum_\alpha \omega_\alpha \hat a_\alpha^\dagger\hat a_\alpha
 \\
          +& \sum_{i=1,2;k\in\{L,R\}}
             \left(V_{ki}^{et}\hat c_k^\dagger\hat d_i e^{-i(-1)^i\omega_0 t/2}
             + H.c.
             \right)
 \nonumber \\
 \label{Vbar}
 \hat{\bar V} =& \sum_{k\neq k'\in\{L,R\}}
        \left(V_{kk'}^{eh}\hat c_k^\dagger\hat c_{k'}\hat d_2^\dagger\hat d_1
              e^{i\omega_0 t}
        + H.c.
        \right)
        \\
        +& \qquad 
        \sum_\alpha\left(V_\alpha^{p}\hat a_\alpha\hat d_2^\dagger\hat d_1
                         e^{i\omega_0 t}
        + H.c.
        \right)
        \nonumber
\end{align}
where
\begin{equation}
 \label{ebar}
 \bar\varepsilon_i = \varepsilon_i - (-1)^i\omega_0/2
\end{equation}

Within rotating wave approximation only diagonal elements of the self-energy
due to coupling to the contacts (electron transfer)
$\Sigma^{et}$, Eq.(\ref{Set}), and self-energy due to coupling to 
electron-hole excitations (energy transfer) $\Sigma^{en}$, Eq.(\ref{Sen}), 
survive
\begin{align}
 \bar\Sigma^{et}_{ii}(\tau_1,\tau_2) =& \Sigma^{et}_{ii}(\tau_1,\tau_2) 
 e^{i(-1)^i\omega_0(t_1-t_2)/2}
 \\
 \bar\Sigma^{en}_{ii}(\tau_1,\tau_2) =& \Sigma^{en}_{ii}(\tau_1,\tau_2)
 e^{i(-1)^i\omega_0(t_1-t_2)}
\end{align}
For self-energy due to coupling to photon bath $\Sigma^p$, Eq.(\ref{Sp}),
we neglect non-diagonal terms, since they contribute to retarded (advanced)
projection only and coupling to the bath is assumed to be small relative
to coupling to the contacts. The self-energy becomes diagonal
\begin{equation}
 \bar\Sigma^{p}_{ii}(\tau_1,\tau_2) = \Sigma^{p}_{ii}(\tau_1,\tau_2)
 e^{i(-1)^i\omega_0(t_1-t_2)}
\end{equation}

Resulting Green functions $\bar G(t_1,t_2)$ depend parametrically on slow 
time variable $t=(t_1+t_2)/2$ through time dependence of the envelope
$\vec F(t)$, Eq.(\ref{H0bar}). Transforming to Wigner coordinates,
taking Fourier transform in the relative coordinate $t_1-t_2$, and 
using gradient expansion,\cite{HaugJauho} leads to the following
expressions for charge 
\begin{align}
 \label{IKbar}
 &\bar I_K(t) = \sum_{n=0}^{\infty}\frac{i^n}{2^n n!}\int_{-\infty}^{+\infty}
 \frac{dE}{2\pi}
 \\
 &\mbox{Tr}\left[
 \frac{\partial^n\mathbf{\bar\Sigma}_K^{<}(E)}{\partial E^n}
 \frac{\partial^n\mathbf{\bar G}^{>}(t,E)}{\partial t^n} -
 \frac{\partial^n\mathbf{\bar\Sigma}_K^{>}(E)}{\partial E^n}
 \frac{\partial^n\mathbf{\bar G}^{<}(t,E)}{\partial t^n}
 \right] 
 \nonumber
\end{align}
and photon 
\begin{align}
 \label{Jabar}
 &\bar J_a(t) = |V_\alpha^p|^2\sum_{n,m=0}^{\infty}\frac{i^{n+m}}{2^{n+m} n! m!}
 \int_{-\infty}^{+\infty}\frac{dE}{2\pi}
 \\
 &\left(\frac{\partial^n}{\partial t^n}\frac{\partial^n}{\partial E^n}
       {\bar G}^{>}_{11}(t,E)\right)
 \left(\frac{\partial^m}{\partial t^m}\frac{\partial^m}{\partial E^m}
       {\bar G}^{<}_{22}(t,E+\omega_\alpha)\right)
 \nonumber
\end{align}
fluxes. Eqs.~(\ref{IKbar}) and (\ref{Jabar}) are main results
of this section. They are to be compared with general expressions
(\ref{IKt}) and (\ref{Jatacc}), respectively.

\section{\label{numres}Numerical results}
We calculate time-dependent transport and optical response by envoking
Runge-Kutta scheme with adaptive stepsize control\cite{recipesC}
to solve numerically system of differential equations (\ref{DysA}).

\begin{figure}[htbp]
\centering\includegraphics[width=\linewidth]{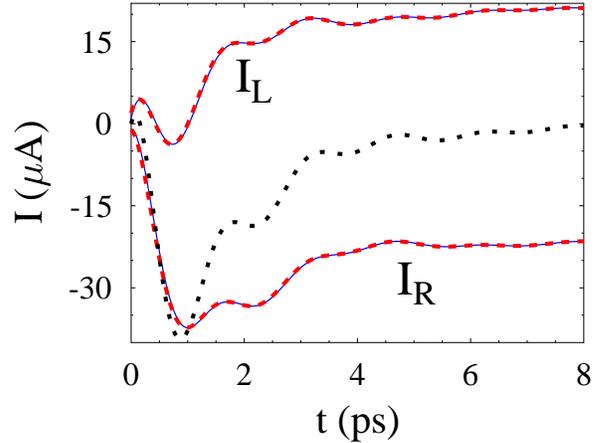}
\caption{\label{fig1}
(Color online) Current on the left, $I_L$, and right, $I_R$,
interfaces vs. time for single level model.
Numerical results (dashed line, red) are compared to analytical expression
(solid line, blue). Also shown is sum  of the currents, $I_L+I_R$ at the two 
interfaces (dotted line, black). See text for parameters.
}
\end{figure}

\begin{figure}[t] 
\centering\includegraphics[width=\linewidth]{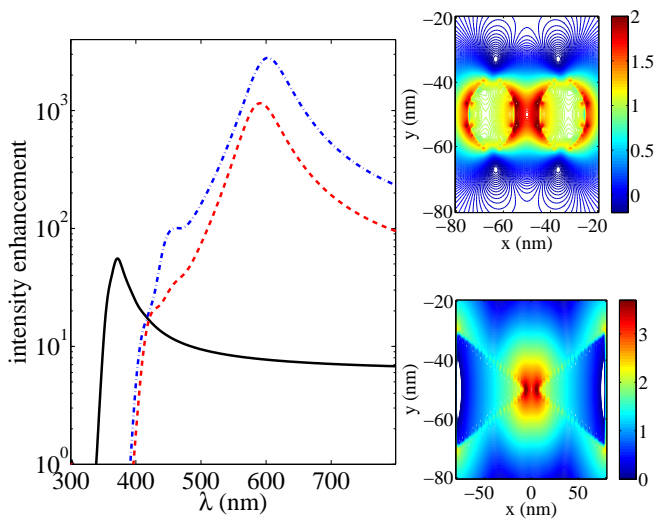}
\caption{\label{fig2}
(Color online) Results of FDTD simulations. Left panel shows intensity 
enhancement as a function of the incident wavelength (in nm) 
in logarithmic scale for two spheres of $20$ nm in diameter with a gap of 
$10$ nm (solid line, black) and bowtie antenna with a gap of $10$ nm (dashed line, red), 
and $5$ nm (dash-dotted line, blue). Top right inset represents steady-state intensity 
enhancement distribution in logarithmic scale for two spheres system 
at the resonant wavelength of $368.202$ nm. 
Lower right inset shows intensity distribution for the bowtie antennas 
with a gap of $5$ nm at $602.647$ nm.
}
\end{figure}

To check accuracy of our numerical approach
we start from a test calculation for a single level model.
Analytical solution is available for the latter.\cite{JauhoWingreenMeir}
In a biased junction ($\mu_L=1$~eV and $\mu_R=-1$~eV) the level is set
below both chemical potentials ($\varepsilon_0=-2$~eV), 
so that initially the level is occupied and current through the junction 
is negligible (escape rates are $\Gamma_L=\Gamma_R=0.2$). At time $t_0$
position of the level is shifted to $0$~eV (steplike modulation).
Here and below we assume Fermi distributions in the leads corresponding to
room temperature $T=300$~K. Figure~\ref{fig1} presents transient current
at the two interfaces (direction from contact into the system is taken
to be positive for both currents) calculated numerically (dashed line) and 
with analytical solution (solid line). Also shown is sum of the currents
at the two interfaces (dotted line). Outflux of electrons from
initially fully populated level into the right contact leads to ringing effect.
Eventually the current achieves steady-state. Our numerical procedure
is seen to give good correspondence with the analytical result.
Below we use similar parameters for calculation of time-dependent
response of the two-level system. 

We consider two geometries of a junction: a bowtie antenna like electrodes
and electrodes in the form of metallic spheres. 
Large single-molecule fluorescence measurements were reported recently
for the former.\cite{bowtie} The latter (molecule between two metallic 
nanoparticles) is customary in experimental setups. 

Both structures are excited by a plane wave polarized along the axis of 
symmetry (i.e. along the axis connecting centers of two spheres, for instance).
The electric field amplitude is then detected as a function of time. 
Recorded amplitudes are Fourier transformed and normalized with respect 
to the incident field amplitude leading to enhancement as a function in the
frequency domain.

Results of our simulations for both geometries are presented in Fig.~\ref{fig2} 
showing intensity enhancements in the main panel. As expected bowtie 
structures result in noticeably higher enhancements reaching $630$ centered 
at $\lambda = 600$ nm for a bowtie antenna with a gap of $5$ nm. 
Two spheres also show significant enhancement of $55$ around 
$\lambda = 370$ nm. We note that the 
bowtie antenna in comparison to two spheres system exhibits two resonances.
The "blue" resonance located at low wavelength
corresponds to rod lightning effect with high enhancement localized primarily 
at the edges of each triangle. This feature disappears from the spectrum once 
sharp corners are replaced with smooth edges.\cite{SukharevSeideman07} 
Top and bottom insets show intensity enhancement distributions
at resonant conditions for the two spheres and bowtie antennas, respectively.  
We place molecular junction in the hot spot regions.

\begin{figure}[t]
\centering\includegraphics[width=\linewidth]{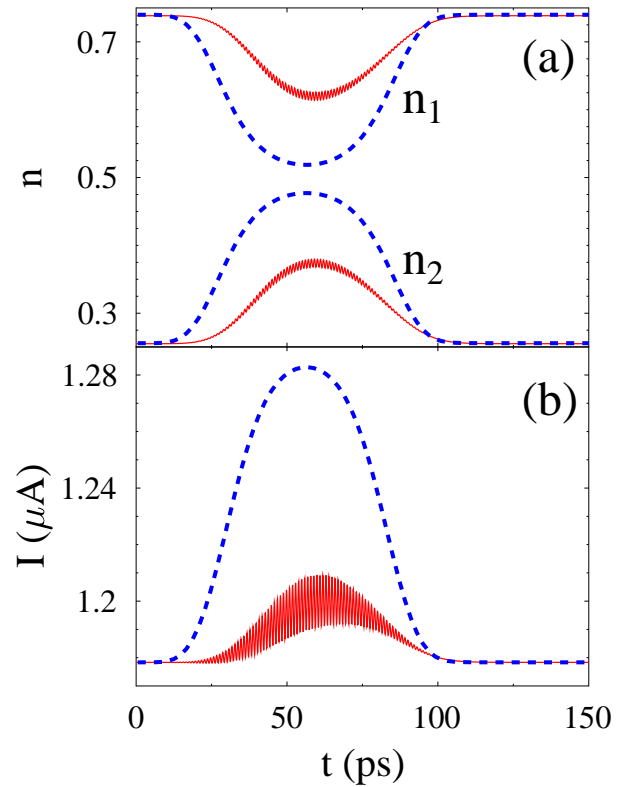}
\caption{\label{fig3}
(Color online) Comparison of exact numerical solution (solid line, red) 
to adiabatic approximation (dashed line, blue) for the two-level (HOMO-LUMO) 
model. Shown are (a) levels populations and (b) current at the left interface
vs. time. See text for parameters.
}
\end{figure}

\begin{figure}[htbp!]
\centering\includegraphics[width=\linewidth]{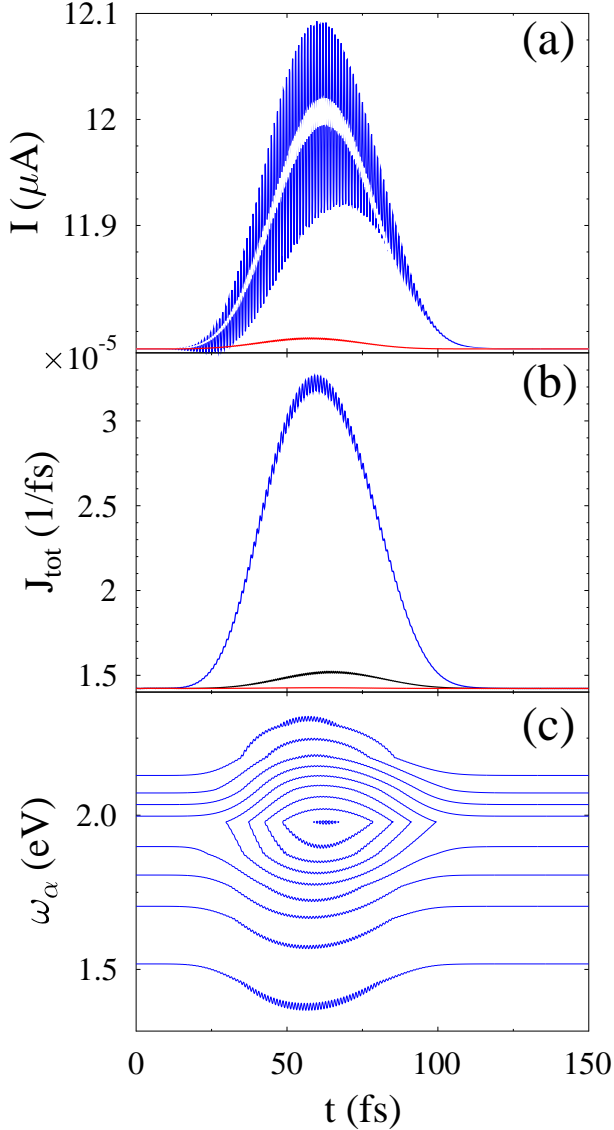}
\caption{\label{fig4}
(Color online) The two-level (HOMO-LUMO) model. 
Shown are (a) current and (b) total optical response, (\ref{Jtot}), 
vs. time for bowtie nanoantennas (the strongest signal, blue) 
and two spheres junction geometries. In the latter case the response is
calculated for two positions of the molecule in the junction: 
in the middle between the spheres (the weakest signal, red)
and closer to one of the spheres (intermediate signal, (a) white silhouette 
and (b) solid line, black). Figure (c) shows contour map of
optical flux, (\ref{Jwa}), for bowtie geometry vs. outgoing frequency and time. 
See text for parameters.
}
\end{figure}

Figure~\ref{fig3}a shows time-dependent populations of molecular
junction driven by external electromagnetic field for the ground, $n_1$,
and excited, $n_2$, states. Time-dependent current at the left interface,
$I_L$, is shown in Fig.~\ref{fig3}b. Parameters of the calculation are
$T=300$~K, $\varepsilon_1=-1$~eV, $\varepsilon_2=1$~eV, 
$[\mathbf{\Gamma}_{K}]_{mm}=0.1$~eV and 
$[\mathbf{\Gamma}_{K}]_{12}=[\mathbf{\Gamma}_{K}]_{21}=0$ ($m=1,2$ and $K=L,R$).
For interaction with electromagnetic field we take
$\vec\mu\,\vec E_0=0.005$~eV, where $\vec E_0$ is amplitude of the external
laser field before enhancement. Bias $V$ is applied symmetrically
$\mu_{L,R}=E_F\pm eV/2$, and the Fermi energy is $E_F=0$.
Results presented in Fig.~\ref{fig3} are obtained for bowtie geometry
with $10$~nm gap at bias $V=2$~V.
Exact numerical calculation (solid line) is compared with adiabatic 
approximation data (dashed line). One sees, that the adiabatic approximation
for realistic parameters provides qualitatively correct results.
It misses however delay (memory) effects and overestimates response signal.
Electromagnetic pulse depletes ground state and populates excited state,
which for the chosen bias leads to increase of current through the junction
due to increase in transmission of the excited state channel 
(see also Fig.~\ref{fig5} below). 

We compare response of the two molecular junction geometries in Fig.~\ref{fig4}.
Bowtie geometry provides stronger local enhancement, and consequently
stronger molecular response. In the case of spherical nanoparticles 
we consider two possible positions of molecule between the electrodes: 
symmetric and asymmetric ($3$~nm shift from the center, where the field
enhancement for the geometry is strongest). These yield weakest and
intermediate signal, respectively. Note, that it is natural to expect that
local field enhancement is stronger for a structure with uneven surface.
Fig.~\ref{fig4}a presents time-dependent current for the three cases.
Total optical response, Eq.(\ref{Jtot}), is shown in Fig.~\ref{fig4}b.
We choose $\eta=5\cdot10^{-5}$ and $\omega_c=2$~eV,
other parameters are as in Fig.~\ref{fig3}. Note much more sensitive character
of optical response to resonant conditions. It results from our choice of
$\gamma_\alpha(\omega)$, Eq.(\ref{gamma}), so that most of the electronic 
excitation contributes to current. While the choice is arbitrary, it
indicates importance of the environment (bath spectral density).
Fig.~\ref{fig4}c shows time-dependent optical spectrum, Eq.(\ref{Jwa}), 
for the bowtie geometry. The signal follows (with a delay) the pulse of the
external field. Asymmetric character of the spectrum relative to resonance,
$\omega_\alpha=2$~eV, stems from overlap of Lorentzians (levels boradening
due to coupling to the contacts) centered on ground and excited states. 

\begin{figure}[htbp]
\centering\includegraphics[width=\linewidth]{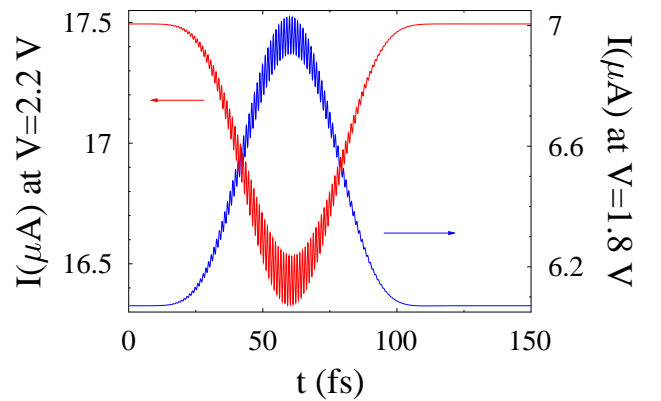}
\caption{\label{fig5}
(Color online) Current vs. time for the two-level (HOMO-LUMO) model
calculated at two different biases. See text for parameters.
}
\end{figure}

Figure~\ref{fig5} shows time-dependent current response to external driving
at two different constant biases. The calculation is done for bowtie geometry
with a gap of $10$~nm,
parameters are the same as in Fig.~\ref{fig3}. For pre-resonant bias, 
$V=1.8$~V, optical excitation is effective in depleting the ground 
and populating the excited states of the molecule, which results in increased
current through both channels. At post-resonant bias, $V=2.2$~V,
the charge transfer channels are open. Here optical excitation contributes 
mostly to decrease in conductance of the ground state and appearance of
leakage current to the left contact in the excited state. This leads to
overall decrease in current through the junction (see also discussion below).

\begin{figure}[htbp]
\centering\includegraphics[width=\linewidth]{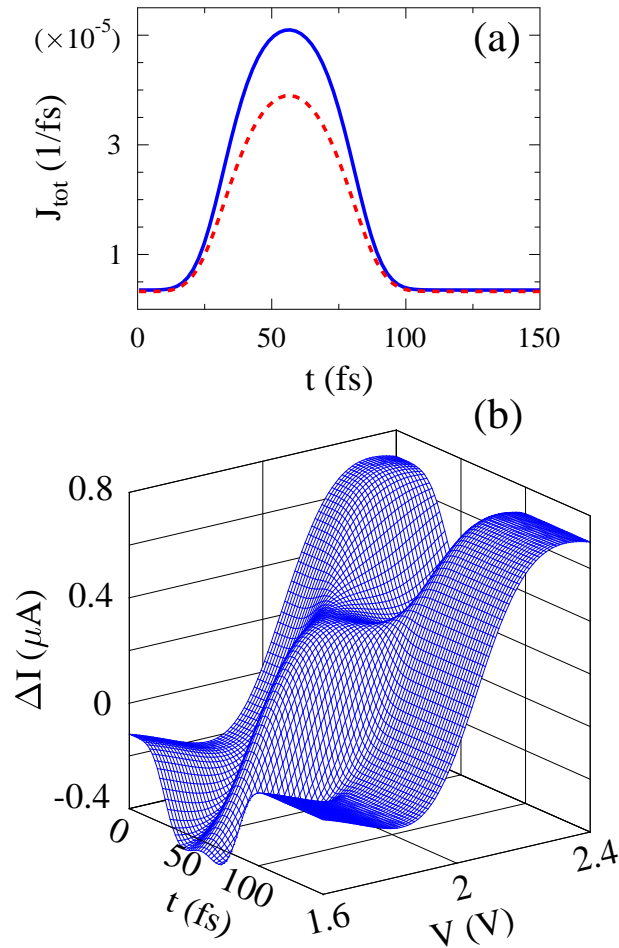}
\caption{\label{fig6}
(Color online) 
Role of energy transfer process. 
Shown are (a) total optical response vs. time with (dotted line, red)
and without (solid line, blue) electron-hole excitations and
(b) difference between current calculated with and without electron-hole
excitations vs. time and bias.
Calculations are performed within adiabatic approximation scheme.
See text for parameters.
}
\end{figure}

Calculations so far disregarded influence of both energy transfer,
Eq.(\ref{Sen}), and external photon bath, Eq.(\ref{Sp}), 
(except its  contribution to optical rate) on electronic distribution 
in the molecule. While the latter can indeed be disregarded due to smallness
of the reasonable coupling parameter (see Ref.~\onlinecite{Nitzan_jcp}
for discussion), the former can make a difference. Here we illustrate
influence of energy transfer process on time-dependent response of the
junction within adiabatic approximation (full numeric calculation
is straightforward but time-consuming). 
Figure~\ref{fig6}a shows total optical response
calculated with (dashed line) and without (solid line) energy transfer
included. Calculation is done for bowtie geometry with $10$~nm gap 
at pre-resonant constant bias $V=1.8$~V. 
Other parameters are as in Fig.~\ref{fig3}.
As expected, energy transfer diminishes optical response of the
junction, since both energy transfer from molecule to contacts and
fluorescence compete for the same excess electronic population in the 
excited state. 
Current change upon including electron-hole excitations into consideration
is more interesting. Interplay between channel blocking and resonant pathways
for electron transfer may lead to increase in current through the junction 
as is illustrated in Fig.~\ref{fig6}b). This effect is similar to
the situation presented in Fig.~\ref{fig5}.

\section{\label{conclude}Conclusion}
We consider a two-level (HOMO-LUMO) model of molecular junction driven 
by external time-dependent laser field. Finite difference time domain 
technique is used to calculate field distribution for two junction 
geometries. Resulting local field at the molecule is considered to be 
the driving force.
We assume that the junction is initially in a nonequilibrium
steady-state resulting from applied constant bias. At time $t_0$
driving force (laser pulse) starts to influence the system. 
Time-dependent transport (charge flux through the junction) and 
optical response (photon flux from the molecule into accepting modes) 
are calculated for a set of geometries and applied biases. We rewrite
a nonequilibrium Green function technique for time-dependent calculation
in a form convenient for treating many-level molecular systems.
Results of the simulations within the approach are compared to approximate
scheme for an adiabatic pumping regime. Note that while our present 
consideration is restricted to driving force applied to the molecule only, 
generalization of the approach to situations of time-dependent bias and/or
coupling between molecule and contacts is straightforward. Extension of
the consideration to realistic molecular devices, taking into account
time-dependent non-equilibrium distribution in the contacts and
spatial profile of the field, and considering 
interplay of time-dependencies of bias and laser field are goals of future 
research. 

\begin{acknowledgments}
M.S. is grateful to ASU financial and technical support (startup funds).
M.G. gratefully acknowledges support by the UCSD (startup funds), 
the UC Academic Senate (research grant), 
and the U.S.-Israel Binational Science Foundation. 
\end{acknowledgments}


\end{document}